\documentclass[twocolumn,showpacs,preprintnumbers,amsmath,amssymb]{revtex4}

\usepackage{graphicx}
\usepackage{dcolumn}
\usepackage{bm}

\begin{document}

\title{Bidirectional wavelength tuning of semiconductor quantum
    dots as artificial atoms in an optical resonator} % \\

% Other possible titles
% OR \\
% Reversible tuning of emission from quantum dots in microtube resonators
%

%\author{S. Mendach}
%\email{s.mendach@physnet.uni-hamburg.de}
%\altaffiliation[Also at]{Physics Department, XYZ University.}%Lines break automatically or can be forced with \\
\author{S. Mendach, S. Kiravittaya,  A. Rastelli, M. Benyoucef, R. Songmuang, and O.G. Schmidt}%
\affiliation{Max-Planck-Institut f\"ur Festk\"orperforschung,
Heisenbergstr. 1, D-70569, Stuttgart, Germany}

\date{\today}% It is always \today, today,

\begin{abstract}
We consider a pair of artificial atoms with different ground state
energies. By means of finite element calculations we predict that
the ground state energies can be tuned into resonance if the
artificial atoms are placed into a flexible ring structure, which
is elastically deformed by an external force. This concept is
experimentally verified by embedding a low density of
self-assembled quantum dots into the wall of a rolled up micro
tube ring resonator. We demonstrate that quantum dots can
elastically be tuned in- and out of resonance with each other or
with the ring resonator modes.
\end{abstract}
% Please remove the PACS description after select them.

\maketitle

%Introduction
The field of solid state cavity quantum electro dynamics (QED) has
gained considerable interest in recent years particularly due to
potential applications in the area of quantum information processing
\cite{Vahala03}. For example, coupling of a single semiconductor
quantum emitter inside a semiconductor cavity has been shown for the
weak \cite{Gerard98,Kiraz01,Pelton02,Badolato05} as well as for the
strong coupling
regime\cite{Reithmaier04,Yoshie04,Peter05,Hennesy07}. At the same
time, great efforts have been made to exert full control over both
spatial~\cite{Schmidt07} as well as spectral
position~\cite{Rastelli07} of more than one quantum emitter.
However, a practical concept to tune two quantum emitters into
resonance with a single cavity mode~\cite{Imamoglu99} has not been
developed for semiconductor systems, yet.\\
Stimulated by recent experiments \cite{Mendach05, Kipp06,
Mendach06, Mendach06b, Songmuang07, Seidl06b}, we here propose to
embed two artificial atoms into the wall of a flexible microtube
optical ring resonator. Our calculations predict that – upon local
deformation of the ring resonator – reversible spectral shifts
into the red and blue of several tens of meV can be achieved for
the two quantum emitters. Depending on the relative position and
magnitude of the applied force the two quantum emitters can be
tuned into mutual resonance and into resonance with the optical
mode. Experimentally, we verify this concept for a low density of
InGaAs quantum dots (QDs) embedded in the wall of a rolled-up
microtube ring resonator. Energy shifts as large as 20 meV are
achieved for the QDs, and optical coincidence between two QDs as
well as a QD and the ring resonator mode are demonstrated. The
feasibility of spatial coincidence with the 3D-confined modes of
the tube resonator~\cite{Mendach06b, Strelow07}, e.g. by placing
quantum dots on patterned holes created by lithography, has
already been demonstrated in other works~\cite{Kiravittaya06} and
is therefore not addressed here.\\
\begin{figure}
\includegraphics[width=8.5cm]{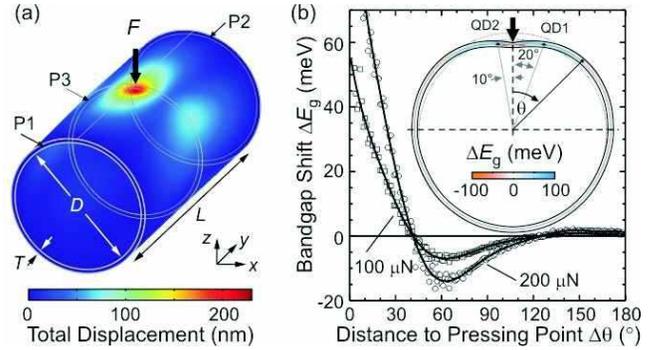}
\caption{(a) 3D schematic and profile of the total displacement
induced by a pressing force $F$. (b) Calculated band gap shift
$\Delta E_{\textrm{g}}$ as function of the azimuthal distance from
the pressing position~($\Delta\theta=\theta-\theta_{P}$) for a force
strength of 100~$\mu$N (open square) and 200~$\mu$N (open circle).
The solid lines are guides to the eye. The inset shows the profile
of the band gap shift in the pressing plane P3 (cf.~(a)) for the
pressing angle $\theta_{P}$~=~0. The positions of QD1 and QD2
correspond to the example discussed in the theory section.}
\end{figure}
Figure~1 shows a point force applied to the top of an optical
micro tube ring resonator. Based on the finite element
method~(FEM) a numerical calculation is performed to
quantitatively describe the tube deformation and the strain
induced energy shifts of two QDs located in the wall of the
deformed tube \cite{comsol}. For our calculations we approximate
the flexible ring resonator by a simple GaAs-tube with inner
diameter $D=$4.3 $\mu$m, tube wall thickness $t=$130 nm and tube
length $L=$10 $\mu$m. Figure~1(a) shows the geometry of the
calculated structure. The tube is fixed at planes P1 and P2 and we
assume isotropic GaAs material parameters. The point force $F$
lies in the pressing plane P3 and is always applied in radial
direction at the outer tube wall surface. The pressing angle
$\theta_{P}$ represents the azimuthal coordinate of $F$, e.g.
$\theta_{P}$ = -10$^{\circ}$ corresponds to a force applied in
radial direction above QD2 while $\theta_{P}$ = 20$^{\circ}$
corresponds to a force applied in radial direction above QD1 (cf.
inset of Fig.~1(b)). The situation shown in Fig.~1(a) and in the
cross section given in the inset of Fig.~1(b) corresponds to
$\theta_{P}$~=~0$^{\circ}$ with an applied force $F$ of 100
$\mu$N. The most prominent feature is a considerable displacement
of the tube wall inwards at the pressing position (red region) and
outwards in two regions approximately $\pm$60$^{\circ}$ away from
the pressing position (light blue on both sides of the tube, only
one is visible in the 3D view). From the displacement we can
derive all strain components induced by the pressing. The strain
components in cartesian coordinates are transformed into polar
coordinates and used to calculate the energy bandgap shift $\Delta
E_{\textrm{g}}$ by applying linear deformation potential theory
\cite{Walle89,Reilly89}. Due to the fact that during pressing the
strain in the tube resonator changes on a large scale compared to
the dimensions of a quantum dot, we do not expect considerable
changes of the quantum dot shape, size and strain profile and
therefore neglect energy shifts connected with these parameters.
The two curves in Fig.~1(b) illustrate the local dependence of the
bandgap shift $\Delta E_{g}$ in the pressing plane P3 for two
different strengths of the applied force (squares $F$=100~$\mu$N,
circles $F$=200~$\mu$N). We clearly see that, depending on the
azimuthal distance from the pressing position
($\Delta\theta=\theta-\theta_{P}$), \textit{both} upward (blue
shift) and downward (red shift) shifts in the range of several
tens of meV are created at the same time for the GaAs band gap. It
is noteworthy that small blue shifts are induced by this pressing
even at positions far from the pressing position ($\Delta\theta$
$>$ 130$^{\circ}$). A false color profile illustration of the
locally varying band gap shift $\Delta E_{g}$ induced in the
pressing plane P3 by a force of 100~$\mu$N is given
in the inset of Fig.~1(b).\\
In order to illustrate the possibility of tuning two QD transition
energies into resonance with a resonator mode we first focus on
one specific example:\\
(i)~QD1 and QD2 are located in the pressing plane P3. (ii)~The
azimuthal coordinate is $\theta$~=~20$^{\circ}$ for QD1 and
$\theta$~=~-10$^{\circ}$ for QD2 [cf. inset of Fig.~1(b)]. Both
QDs are located in the middle of the tube wall. (iii)~QD1 is
assumed to emit at 925 nm ($E_{\textrm{QD1}}$ = 1.3405 eV). The
emission energy of QD2 is assumed to have a slightly higher value
(910 nm: 1.3626 eV). (iv)~The optical mode emission line of the
microtube resonator is located in between at 914.5 nm (1.3560 eV).
(v)~The mode line does not shift when a force is applied to the
tube. Pressing experiments show that shifts due to a change in the
resonator shape or due to strain induced changes in refractive
index are small and can be neglected. (vi)~Tilting of the energy
band edges due to the radial strain gradient and a change of
quantization effects due to the pressing are neglected.
\begin{figure}
\includegraphics[width=8.5cm]{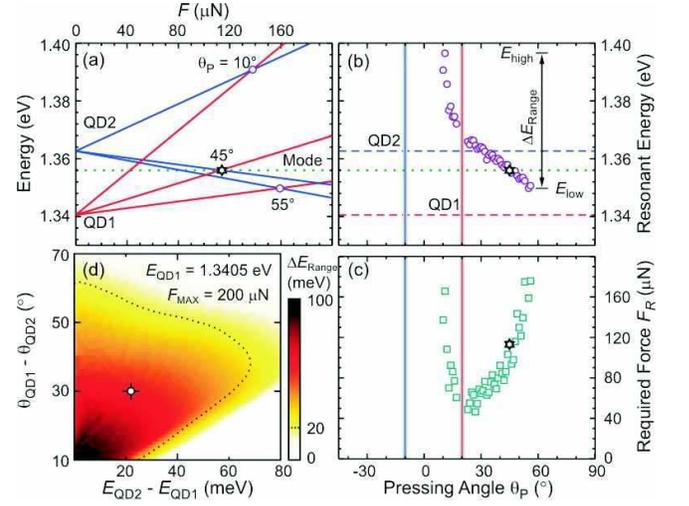}
\caption{The stars in (a), (b), (c) indicate the same QD-QD-Mode
resonance. (a)~Calculated transition energies of QD1 and QD2 as a
function of applied force $F$. The optical mode energy is presented
as a dotted line. Open circles indicate QD-QD resonance.
(b)~Resonant energies as a function of the pressing angle. Initial
transition energies and azimuthal positions of QD1 and QD2 are shown
as lines. (c)~Required force $F_{R}$ needed to bring QD1 and QD2
into resonance at the resonant energies shown in (b). (d)~Width of
resonant energy range (cf.~(b)) as a function of the azimuthal and
spectral distance of QD1 and QD2. The dotted line marks the area
with $\Delta E_{range}$~$>$~$\Delta E_{Modes}$, detailed
explanations see text.}
\end{figure}
Figure~2(a) illustrates the emission energy of QD1 (red lines) and
QD2 (blue lines) for three different pressing angles as a function
of the pressing force $F$. For all pressing angles we find a force
strength which leads to a resonance of QD1 and QD2 (intersection of
the lines marked with a circle/star and $\theta_{P}$). In general,
resonance is only possible within a certain energy range~$\{E_{low},
E_{high}\}$ which corresponds to a certain set of pairs~($F$,
$\theta_{P}$). As shown in Fig.~2(b) the spectral position of the
resonance between QD1 and QD2 is a continuous function of the
pressing angle within this energy range. This means that QD1 and QD2
can be brought into resonance at any energy within this energy
range, or in other words, it is possible to bring QD1 and QD2 into
resonance with any resonator mode line within this energy range.
Interestingly, the width of the energy range $\Delta
E_{Range}$~=~$E_{high}-E_{low}$~$\approx$~47~meV shown in Fig.~2(b),
which is derived for moderate force strengths of up to
$F$~$=$~200~$\mu$N (cf. Fig.~2(c)), is two times larger than the
typical distance between two resonator modes ($\Delta
E_{Modes}$~$\approx$~20~meV, cf. experimental part). Therefore,
there is always at least one mode line reachable for a resonance
with QD1 and QD2. For the mode line energy assumed in this example
(dotted line in Fig.~2(a) and~(b)) we obtain resonance with QD1 and
QD2 for $\theta_{P}$~=~45$^{\circ}$, $F$~=~113~$\mu$N. This
resonance is marked by the stars in Figs.~2(a)-(c).\\
Finally, we consider the case where emission and location of QD2
is varied while QD1 is fixed at 1.3405 eV and
$\theta$~=~20$^{\circ}$. The false color plot in Fig.~2(d) shows
the width of the resonant energy range as a function of the
spectral and local (azimuthal) distance of QD1 and QD2. In the
white area the QDs are spectrally and locally far apart and no
resonance is possible. For QDs both spectrally and spatially
closer together large resonant energy ranges of up to
approximately~100~meV are possible (dark area). The dotted line
marks the area with $\Delta E_{Range}$~$>$~$\Delta E_{Modes}$,
i.e. in this area resonance of QD1, QD2 and a resonator mode is
always possible. Most situations in an experiment are likely to
occur in this area, which underlines the relevance of our
approach. The example discussed above (condition (i)-(vi)) is
marked by the white dot in Fig.~2(d).\\
To experimentally verify the feasibility of the above tuning
concept we proceed as follows: The flexible tube resonators are
fabricated by rolling-up strained semiconductor bi-layers grown by
molecular beam epitaxy (MBE)~\cite{Prinz00, Schmidt01}. Recently,
whispering gallery mode-like resonances in such structures could
be demonstrated by using high density self-assembled
QDs~\cite{Kipp06, Mendach06b} or high density Silicon nano
clusters~\cite{Songmuang07} as inner light sources. Here we
introduce the possibility to perform single QD spectroscopy in
rolled-up micro tube resonators by embedding low density InAs-QDs
into the resonator walls. For that purpose, we used the following
MBE-layer sequence: 400~nm GaAs buffer layer, 20~nm
AlAs~sacrificial layer, 15~nm In$_{17}$Al$_{20}$Ga$_{63}$As
strained layer, 15~nm GaAs, nominal 1.8 ML InAs, and 25~nm GaAs
capping layer. An \emph{in-situ} partial capping and annealing
step~[21] was used to tune the initial emission wavelength of the
QDs into the sensitivity range of the Si-detector used in the
photoluminescence measurements described below. After MBE-growth,
the wafer is exposed to a three step lithographic procedure to
obtain arrays of suspended micro tube resonators:
Step~1:~Definition of U-shaped strained mesa. Step~2:~Definition
of starting edges. Step~3:~Rolling-up of the strained layers by
selectively etching the AlAs layer with HF. More details of the
procedure can be found in Ref.~\cite{Kipp06}. An optical
microscope image of a typical micro tube resonator is shown in
Fig.~3(a). The resonator was rolled-up from the U-shaped area
bordered by the dotted line. The suspended state of the resonator,
obtained by the higher number of rotations in the bearings, is
necessary to avoid leakage of the modes into the GaAs-substrate.
\begin{figure}
\includegraphics[width=8.5cm]{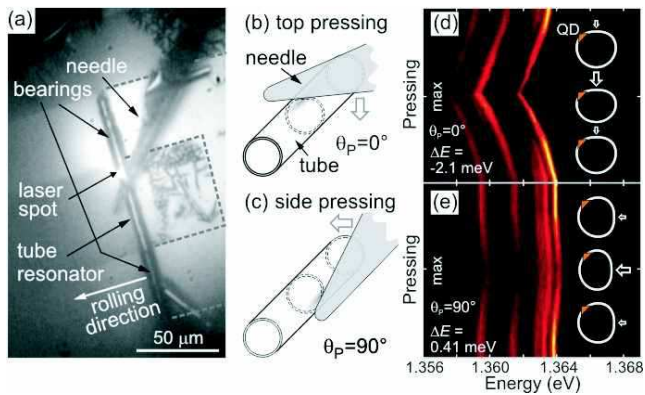}
\caption{(a) Optical microscopy insight into the cryostat: A rolled
up micro tube resonator with a glass needle used for in-situ
pressing. (b),(c)~Schematic of the resonator and the glass needle
for top- ($\theta_{P}$~=~0$^{\circ}$) and side-
($\theta_{P}$~=~90$^{\circ}$) pressing. (d),(e)~Evolution of the PL
from a single QD during pressing for $\theta_{P}$~=~0$^{\circ}$ and
$\theta_{P}$~=~90$^{\circ}$. The inset shows a cross-sectional
schematic of the deformed tube. The triangle indicates the azimuthal
position of the QD.}
\end{figure}
\begin{figure}
\includegraphics[width=8.5cm]{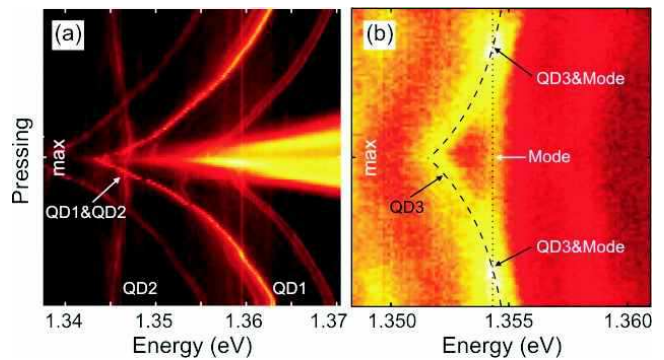}
\caption{(a)~Evolution of PL from two QDs during pressing with
$\theta_{P}$~=~0$^{\circ}$. As indicated by the arrow, resonance of
QD1 and QD2 occurs at 1.3472~eV. (b)~PL of a single QD and a
resonator mode. The mode~(dotted line) is almost independent of the
applied pressing force and is crossed two times by the QD line~(QD3,
dashed line) which reversibly shifts with the applied force.}
\end{figure}
After preparation, the sample is mounted in a cold-finger helium
flow cryostat which can be moved by computer controlled xy-linear
translation stages for exact positioning with a spatial resolution
of 50 nm. Micro photoluminescence ($\mu$-PL) measurements are
performed at $T=8K$ using a frequency-doubled Nd:YVO4-laser
operating at 532 nm. The laser is focused by a microscope
objective (with numerical aperture NA = 0.42) to a spot diameter
of 1.5 $\mu$m. The same microscope objective is used to collect
the PL emission. The collected luminescence is then spectrally
filtered by a monochromator equipped with a liquid nitrogen cooled
charge coupled device (CCD). To in-situ apply forces to the tube
resonators while recording the change in PL we employ a glass
needle mounted on a x-y-z piezo translation stage (Attocube
Systems). Except for quantum dots in the direct vicinity of the
pressing point the finite sized glass needle well-resembles the
point force used in our calculations.\\
After selecting an adequate resonator, we focus on the QD or pair
of QDs to be investigated and position the glass needle next to
the laser spot. The orientation of the glass needle relative to
the micro tube resonator is illustrated in Fig.~3(a)-(c). Pressing
on top of the tube (cf. Fig.~3(b)) corresponds to
$\theta_{P}$~=~0$^{\circ}$ and pressing from the side
(cf.~Fig.~3(c)) corresponds to $\theta_{P}$~=~90$^{\circ}$. These
two pressing angles are used in our experiment to record
the PL as a function of the pressing force.\\
First we demonstrate the bi-directional tuning of a QD:
Figure~3(d) and (e) show the evolution of the PL of a single QD in
a tube resonator during pressing. We did not attempt to quantify
the actual force applied to the tube. The insets illustrate the
estimated tube deformation corresponding to
$\theta_{P}$~=~0$^{\circ}$ and $\theta_{P}$~=~90$^{\circ}$. The
diameter and the tube wall thickness of the rolled-up tube
resonators correspond to the values used in the above calculations
($D$=4.3 $\mu$m, 55~nm strained mesa rolled-up in 2.3~rotations
result in $t\approx$130~nm overall tube wall thickness). As
predicted, both reversible blue- and red shifts of the whole
emission spectrum are observed. As the focus of the objective lens
of the $\mu$-PL setup is optimized on the QD before pressing, we
slightly loose excitation intensity and collection efficiency
during pressing. This effect can be turned around by optimizing
the focus on the QD when the resonator is pressed (not shown). For
$\theta_{P}$~=~0$^{\circ}$ (Fig.~3(d)) the QD spectrum red-shifts
by 2.1~meV and turns back to the initial position as soon as the
force is released. For $\theta_{P}$~=~90$^{\circ}$ (Fig.~3(e)) we
find a reversible blue-shift of 0.41~meV. From the spectral shifts
we estimate the azimuthal position of the QD to be at
$\theta\sim-45^{\circ}$ (see insets of Fig.~3(d) and (e)). This
value agrees well with the
laser spot position optimized for the QD (cf. Fig.~3(a)).\\
Figure~4(a) shows the simultaneous bi-directional tuning of two
different QDs. In this case, the collection is optimized for the
emission of QD1 and the pressing angle is $\theta_{P}=0^{\circ}$.
Two groups of lines, which shift differently, can be attributed to
the emission of QD1 and QD2, respectively. A crossing of two lines
is obtained at 1.3472~eV (see arrow), which indicates spectral
resonance of the two QDs. In this case, QD1 shows a large red
shift of $\sim$20 meV while QD2 shows a blue shift of $\sim$3 meV.
Interestingly, the spectral lines of QD1 change their relative
distance during the pressing process. This cannot be understood
within our model, which considers only the bulk band gap shift. It
might rather be explained by a change of the QD wave functions
shape induced by anisotropic stress~\cite{Stier99}. After
releasing the pressing force, the spectrum returns to the initial
state. It is noteworthy that the $\theta$-dependent strain state
of the deformed tubes (cf.~Fig.~1(b)) also causes a splitting of
the InAlGaAs related emission line, which occurs as soon as the
tubes are deformed (not shown).\\
Finally, Fig.~4(b) illustrates the resonance of a single QD with a
tube resonator mode. The mode line spacings in our resonators are
typically in the order of 20~meV. The quality factor ranges
between 1000 and 4000. QD3~(dashed line) can easily be tuned in
and out of resonance with the mode~(dotted line) by changing the
applied force strength (pressing angle
$\theta_{P}$~=~0$^{\circ}$). The resonance occurs at 1.3545 eV.
While the QD shows a redshift of $\sim$2.5 meV, the optical mode
remains constant within $0.2~$meV.\\
In conclusion, controlling the strain state of semiconductor QDs
in a resonator offers a promising route towards mode mediated
resonance of two artificial atoms. By means of FEM simulations we
predict that spectral coincidence can be achieved if the QDs are
distributed on a ring in a flexible tube resonator. Deforming this
tube resonator in an adequate way can be used to induce QD-QD,
QD-mode or QD-QD-mode resonances. First experiments demonstrate
spectral shifts of up to 20~meV, bi-directional tuning of a single
QD, simultaneous bi-directional tuning of two QDs, as well as
spectral coincidence between two QDs and between a single QD and a
resonator mode. Our technique to achieve spectral coincidence
might be combined with existing techniques for spatial
coincidence~\cite{Kiravittaya06} to elaborate the exciting field
of semiconductor based quantum electrodynamics in a deterministic
fashion.

This work was financially supported by BMBF (Contract No. 03N8711)
and the DFG (Contract No. Schm1298/5-1).

\bibliography{manuscript}
\end{document}